\documentclass[prd, twocolumn, nofootinbib, floatfix]{revtex4-1}

\usepackage[mathscr]{euscript}
\usepackage{amsmath}
\usepackage{graphicx}
\usepackage{dcolumn}
\usepackage{bm}
\usepackage{epsfig}
\usepackage{amssymb,latexsym,mathrsfs}
\usepackage{graphicx}
\usepackage{color}
\usepackage{hyperref}
\usepackage{float}
\usepackage{diagbox}

\definecolor{vividviolet}{rgb}{0.62, 0.0, 1.0}
\definecolor{amaranth}{rgb}{0.9, 0.17, 0.31}
\definecolor{palatinateblue}{rgb}{0.15, 0.23, 0.89}
\definecolor{brightpink}{rgb}{1.0, 0.0, 0.5}
\definecolor{cornflowerblue}{rgb}{0.39, 0.58, 0.93}
\definecolor{deepcarminepink}{rgb}{0.94, 0.19, 0.22}
\definecolor{radicalred}{rgb}{1.0, 0.21, 0.37}

\hypersetup{ linktoc=all,
    colorlinks, linkcolor={palatinateblue},
    citecolor={brightpink}, urlcolor={amaranth}
}
\newcommand{\changeurlcolor}[1]{\hypersetup{urlcolor=#1}}
\newcommand{\be}{\begin{equation}}
\newcommand{\ee}{\end{equation}}
\newcommand{\bs}{\begin{split}} 
\newcommand{\bea}{\begin{eqnarray}}
\newcommand{\eea}{\end{eqnarray}}

\begin{document}

\title{Extreme Hawking Radiation}
\author{Michael R.R. Good}
\affiliation{Physics Department \& Energetic Cosmos Laboratory, \\
Nazarbayev University, Nur-Sultan, Kazakhstan
}

\begin{abstract} 
 Modeling the collapse of an extreme Reissner-Nordstr\"om (ERN) black hole by solving the corresponding moving mirror model for the trajectory that asymptotically approaches uniform acceleration, we obtain the non-zero beta coefficients for all times.   Finite energy is emitted, the radiation spectra is non-thermal (non-steady / not Planck), soft particles characterize the evaporation, and particle production at ultra-late times is damped.  Entanglement entropy diverges with no Page curve turn-over, demonstrating non-thermal information loss. The radiation obeys time-reversal symmetry. 
\end{abstract} 

\date{\today} 

\maketitle
\section{Introduction} 
The extreme Reissner-Nordstr\"om (ERN) black hole has played an important and fascinating role in fundamental theoretical physics.  It is the only static spherically symmetric black hole solution to Einstein’s equations with zero surface gravity (i.e. undefined temperature  \cite{LRS, Anderson:2000pg}), and has been central to our nascent understanding of the non-thermal {\it{classical}} nature of gravitational systems.  Furthermore, since extreme black holes have been crucial for development of a statistical origin of black hole entropy \cite{Strominger:1996sh}, their study has been fruitful for an understanding of the {\it{quantum}} nature of gravitation.

On the experimental side, there is the dynamical Casimir effect which has empirical confirmation (see \cite{Dodonov:2020eto} for a brief timely review and numerous references therein), promoting the 50 year old moving mirror model \cite{Moore}, to a laboratory demonstration of spontaneous emission of particles generated by an accelerated boundary condition.  While the single moving mirror model \cite{DeWitt, Davies1, Davies2} has a long history as an analog system for understanding late-time Hawking radiation \cite{Hawking}, it continues to provide new insights into the equivalence principle's impact on the emission of particles in quantum theory \cite{pagefulling,SWP} and the related Unruh effect \cite{Unruh}.

The equivalence principle links both the collapse to an ERN black hole and an asymptotically uniformly accelerating moving mirror \cite{LRS} because the main property of the incipient ERN is its zero surface gravity. 


Einstein's happy thought tells us that an observer asymptotically moving along with the mirror will feel constant proper acceleration.  This is similar to the force of gravity upwards felt by a standing observer on the Earth.  Vice versa, the curved spacetime of the ERN black hole results in zero acceleration of a static observer near the horizon as measured at infinity. 

\subsection*{Early-times vs. Late-times}

The equivalence happens asymptotically at late times, and so our question is: {\it{what happens at early times?}}  This is where the literal information content concerning collapse, as contained in the `extreme' radiation, is located. Twenty years ago Liberati-Rothman-Sonego (LRS) found the late time behavior \cite{LRS, Liberati:2000zz} of the spectrum of the ERN black hole. That is, they were the first to see that an incipient extreme black hole is modeled at late times by a uniformly accelerated
mirror. Our results provide a focus on the early times by solving for the full all-time spectrum.

More specifically, the beta coefficients found by LRS are those of the uniformly accelerated mirror (e.g. \cite{Davies2, B&D}), and thus, only reveal that soft particles are created at some time in the late stages of collapse, which does not necessarily mean that particles carrying energy are emitted at late times.  In fact, the results of LRS demand there is no energy radiated at late times.  The soft particle production by itself does not contain the energy or information relevant to details of collapse, nor does it mean that particle creation takes place at a steady rate. We shall demonstrate the creation rate of all particles at all times, extracting information about the collapse.  We do this by time evolving the particle creation with wave packet localization. 
\subsection*{No-hair theorem and CCC}
The spectra resulting from the 1+1 dimensional mirror’s worldline is the same as that of the ERN black hole, up to gray-body factors due to the unique $ds^2$ of ERN spacetime and due to the extra space dimensions of 3+1 geometry. LRS concluded that incipient extreme RN black holes create particles with a non-thermal spectrum, and we demonstrate that the full non-asymptotic solution does indeed possess a non-thermal spectrum for the ERN black hole, even during early collapse.  With an exact analytic form of the beta coefficients, the spectrum for all times reveals the particle production evolution leading to the damping at late stages of collapse, concluding that the no-hair theorem is preserved because there is no way to extract information from infinite soft particles at late-times.

One should suspect the cosmic censorship conjecture (CCC) to be violated because neutral scalar particles evaporating leads to a lessening of mass, $M$, all the while the star maintains a fixed charged $Q$, eventually leading to a flippening where the overbalance $Q^2 > M^2$ results in a naked singularity.   However, we shall see that asymptotically the energy flux is zero, and only a finite amount of energy is evaporated, so that an old ERN black hole does not lose any more mass, despite the emission of particles. The CCC is actually preserved by the quantized fields \cite{Hiscock}, even though infinite particles (soft) are radiated.  The results answer the question of how, in the presence of quantum radiation, the formation of a naked singularity is prevented.

This paper is organized as follows:  Section \ref{sec:ERN} contains the ERN metric and matching condition for collapse.  Section \ref{sec:Traj} reveals the moving mirror, and computes the dynamics resulting in asymptotic uniform acceleration.  Section \ref{sec:EE} motivates the use of entanglement entropy to describe 1+1 dimensional black holes and moving mirrors, deriving the entanglement-rapidity formula, connecting it to thermodynamic entropy.  In Section \ref{sec:stress} we compute the stress tensor and total evaporation energy.  Section \ref{sec:particles} demonstrates the all-time spectrum, particle count, a symmetry in time, and compares with the eternally uniform accelerated mirror.  In Section \ref{sec:DISC} we discuss limitations and extensions of the model, and in Section \ref{sec:conc}, we conclude.  Units are $G = \hbar = c  = 1$.

\section{Extreme Reissner-Nordstr\"om} \label{sec:ERN}
\begin{figure}[ht]
\centering 
\includegraphics[width=3.2in,height=3.2in,keepaspectratio]{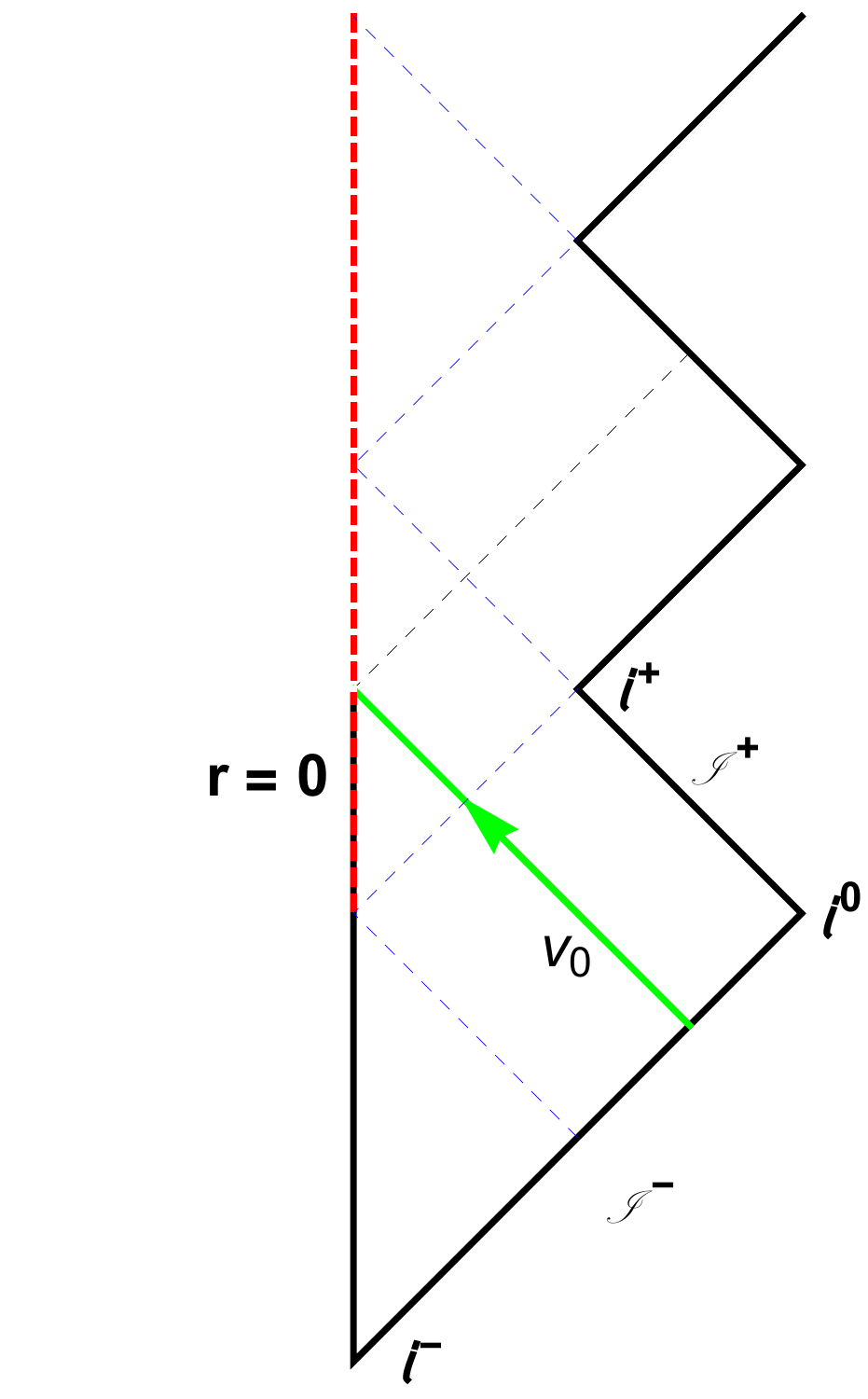} 
\caption{
A Penrose conformal diagram of a collapse of a null shell to form an extreme Reissner-Nordstr\"om black hole. The red dashed line is the irremovable time-like singularity.  The black-red-dashed line is the position at $r=0$ where modes pass through $r=0$ but get lost nevertheless, never reaching $\mathscr{I}^+$.  The thin dashed black line marks the last null ray that passes through $r=0$ without hitting the singularity.  The green line is the null shell, $v_0$.  See Balibinot et al. \cite{Balbinot:2004jx}, Fabbri-Navarro-Salas  \cite{Fabbri} and Carter \cite{Carter} for illustrative ERN diagrams.}
\label{fig:ERNpenrose} 
\end{figure}

\begin{figure}[ht]
\centering 
\includegraphics[width=2.4in,keepaspectratio]{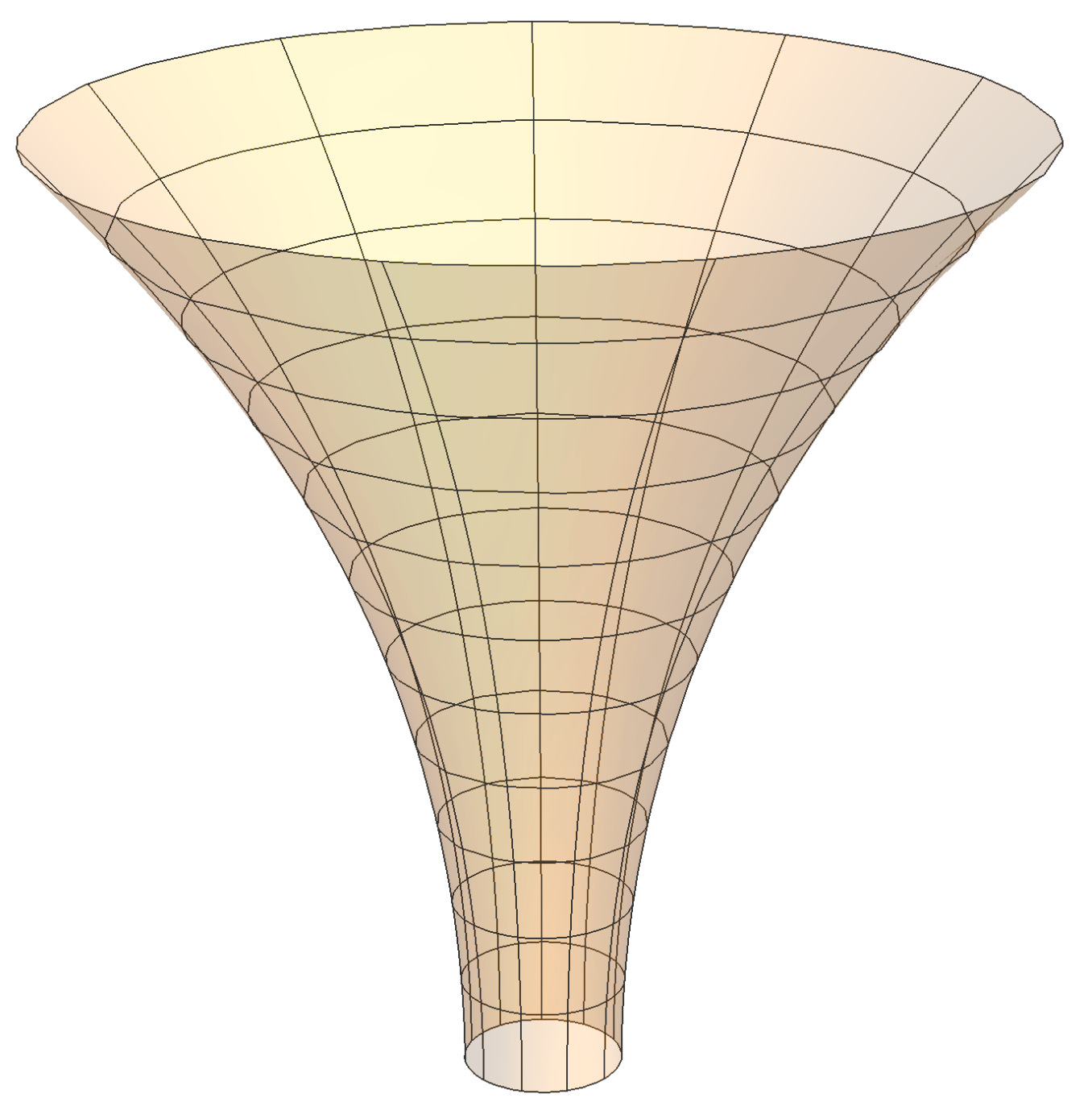} 
\caption{
Infinite throat, $r\rightarrow M$, with an asymptotically flat region for the spatial geometry of the ERN metric, Eq.~(\ref{metric}).  Setting $r=M(1+\epsilon)$, where $\epsilon$ is small, $ds^2 \approx -\epsilon^2 dt^2+(M^2/\epsilon^2) d\epsilon^2 + M^2d\Omega^2$.  The first two terms are $\textrm{AdS}_2$, the last term is $\textrm{S}^2$.   The spacelike distance between an outside point and $r=M$ horizon is infinite as computed by the proper length, $L=\int f^{-1/2} dr$. Time-like curves (and null-rays) reach the horizon crossing the throat in a finite affine time. Evaluated at the horizon, Eq.~(\ref{metric}), has zero surface gravity: `float in the throat'. }
\label{fig:throat} 
\end{figure}

The outside metric of the ERN collapse system, see Fig.~(\ref{fig:ERNpenrose}), is given by the ERN geometry,
\be ds^2 = -\left(1-\frac{M}{r}\right)^2 dt^2 + \left(1-\frac{M}{r}\right)^{-2} dr^2 + r^2 d\Omega^2, \label{metric} \ee
which is most simply interpreted as the exterior field of a spherically charged dust cloud in equilibrium between gravitational attraction and electrostatic repulsion \cite{Carter}. Using the double null coordinate system $(u,v)$, and $u = t-r^*$, and $v = t+r^*$, with the appropriate tortoise coordinate \cite{Fabbri}, 
\be r^* = r + 2M\ln\frac{|r-M|}{M}-\frac{M^2}{r-M}.\ee
one has the metric for the outside collapse geometry,
\be ds^2 =-f(r) du dv + r^2 d\Omega^2,\ee
 where 
\be f(r) = \left(1-\frac{M}{r}\right)^2.\ee
As can be seen, the horizon is at $r=M$, and near it, spacetime looks like an infinite throat, see Fig.~(\ref{fig:throat}).  The surface gravity is calculated to be $\left.f'(r)\right|_{r=M} = 0$. Perhaps then, with this special `equilibrium', one would be led to think there is a temperature and that collapse may evolve to absolute zero, but what then is the spectrum of radiation during collapse?  As is known, thermodynamics is not at play here (or at least categorically different), and we will show explicitly, the all-time spectrum is not Planckian. Even though the system temperature is left undefined, it is possible to find the non-thermal extreme Hawking radiation spectra. 


The matching condition (see e.g. \cite{Balbinot:2004jx,purity,Fabbri,LRS}) between the flat inside geometry, described by inside coordinates, $U=T-r$, and $V=T+r$, is the trajectory of the incipient black hole origin, expressed in terms of the outside function, $u(U)$, dependent on inside coordinate, $U$:
\be u(U) = U+\frac{4M^2}{ \left(v_H-U\right)}-4M \ln \left(\frac{v_H-U}{2M}\right).\ee
The matching happens along the shell, $v_0$, where $v_0 -v_H\equiv 2M$.  Regularity of the field at $r=0$ ensures the form of field modes, such that $U\leftrightarrow v$ identification can be made for the red-shifting right movers.  The mirror trajectory, $f(v) \leftrightarrow u(U)$ is then known, which we examine in the next section in the more simple background of flat space.\footnote{The geometry is not technically Minkowski because of the presence of the mirror.}

\section{Trajectory and Dynamics}\label{sec:Traj}
\begin{figure}[ht]
\centering 
\includegraphics[width=3.2in]{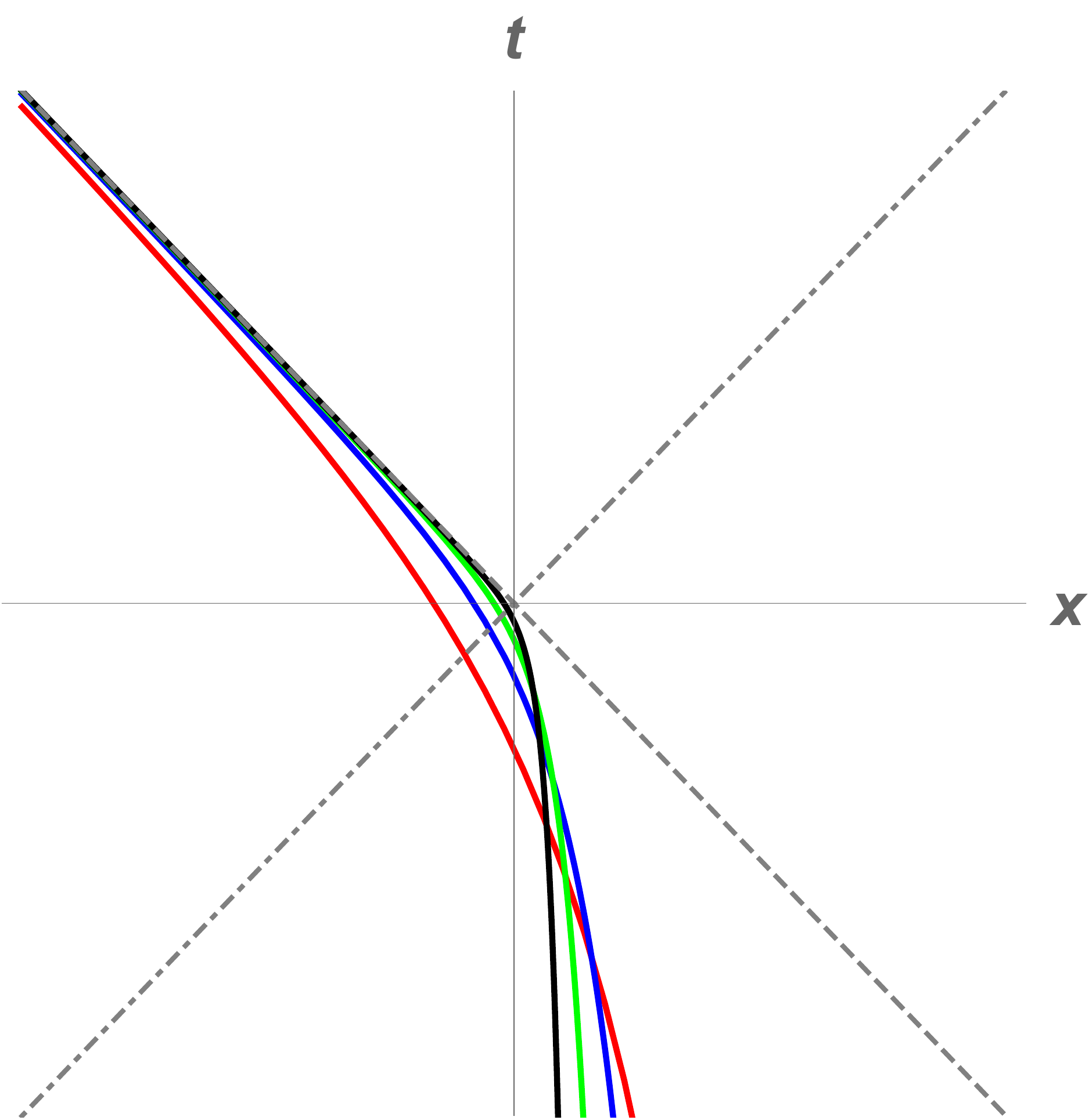} 
\caption{The trajectory Eq.~(\ref{t(x)}), in a spacetime plot, with different scaling for asymptotic future proper acceleration, $\kappa$.  The red, blue, green, black curves correspond to $\kappa = 1/2, 1, 2, 4$, respectively.  The horizon has been set to $v_H = 0$ for all the motions. The mirror starts asymptotically static, but has uniform acceleration in the far future.  
}
\label{fig:SpacetimePlot} 
\end{figure}

\begin{figure}[ht]
\centering 
\includegraphics[width=3.2in]{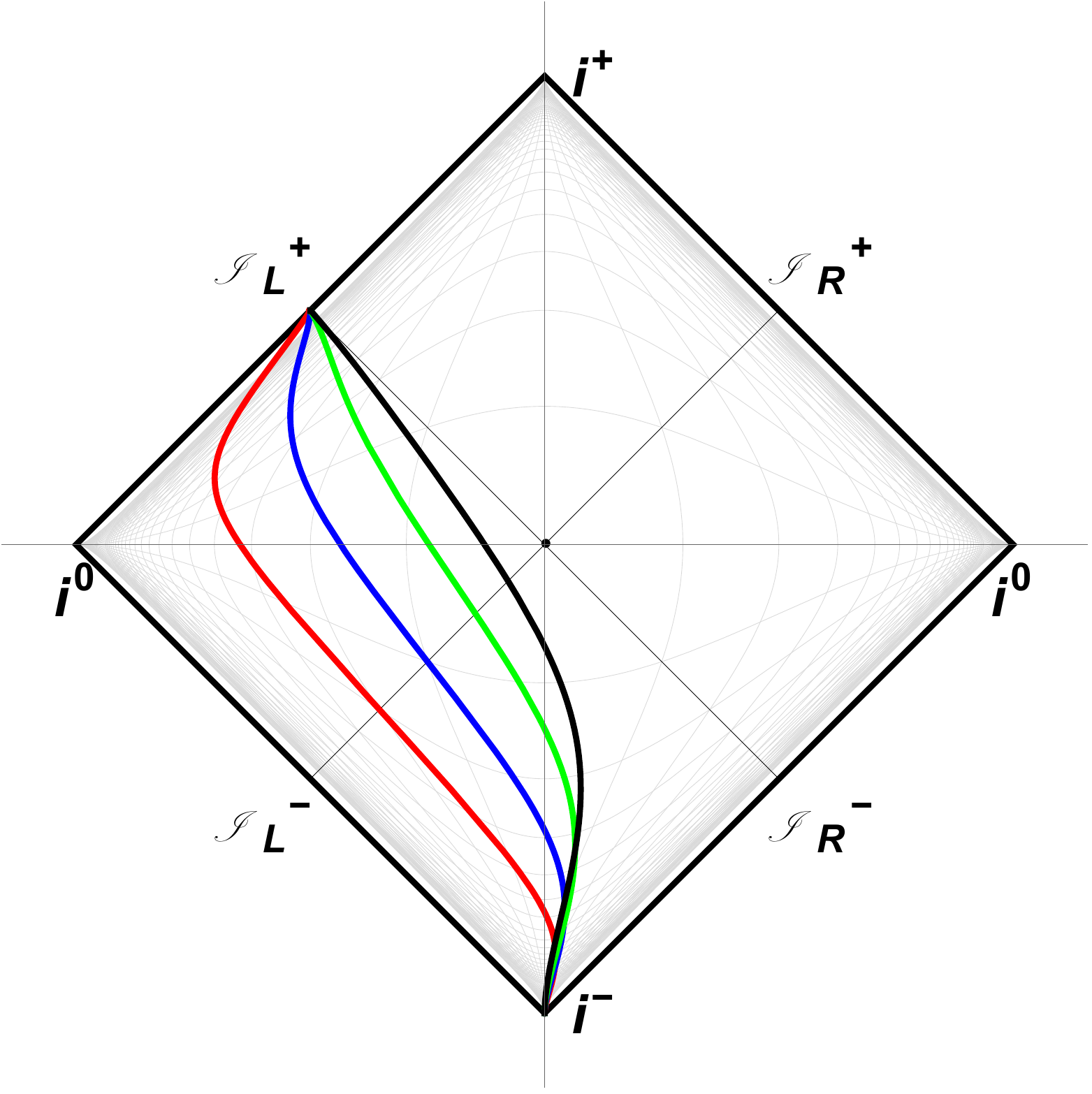} 
\caption{The trajectory Eq.~(\ref{t(x)}), in a Penrose conformal diagram, with different scaling for asymptotic future proper acceleration, $\kappa$.  The red, blue, green, black curves correspond to $\kappa = 1/2, 1, 2, 4$, respectively. This is the same color scheme of Fig~(\ref{fig:SpacetimePlot}).  The horizon has been set to $v_H = 0$ for all the motions.  
}
\label{fig:PenrosePlot} 
\end{figure}

For a massless scalar field in $1+1$ flat spacetime, the corresponding ERN moving mirror  trajectory, where $\kappa \equiv 1/(2M)$, and $\kappa(v_0-v_H) \equiv 1$, is 
\be f(v) = v+\frac{1}{\kappa ^2 \left(v_H-v\right)}-\frac{2}{\kappa} \ln \left(\kappa  \left(v_H-v\right)\right),\label{f(v)}\ee
expressed in null coordinates $(u,v)$ with $u$-function, $f(v)$, as a function of null coordinate advanced time $v$.  The inverse is the usual advanced time $v$-function, $p(u)$, which is a function of retarded time $u$.  Perhaps more intuitively, we express this motion in spacetime coordinates as a time, $t(x)$, function of coordinate space, $x$,
\be t(x) = v_H-\frac{1}{2 \kappa  W\left(\frac{e^{-\kappa  x}}{2}\right)}-x.\label{t(x)} \ee
The trajectory in spacetime coordinates is plotted in a spacetime plot Fig.~(\ref{fig:SpacetimePlot}).  A conformal diagram of the accelerated boundary is given in Fig.~(\ref{fig:PenrosePlot}).  

\begin{figure}[ht]
\centering 
\includegraphics[width=3.2in]{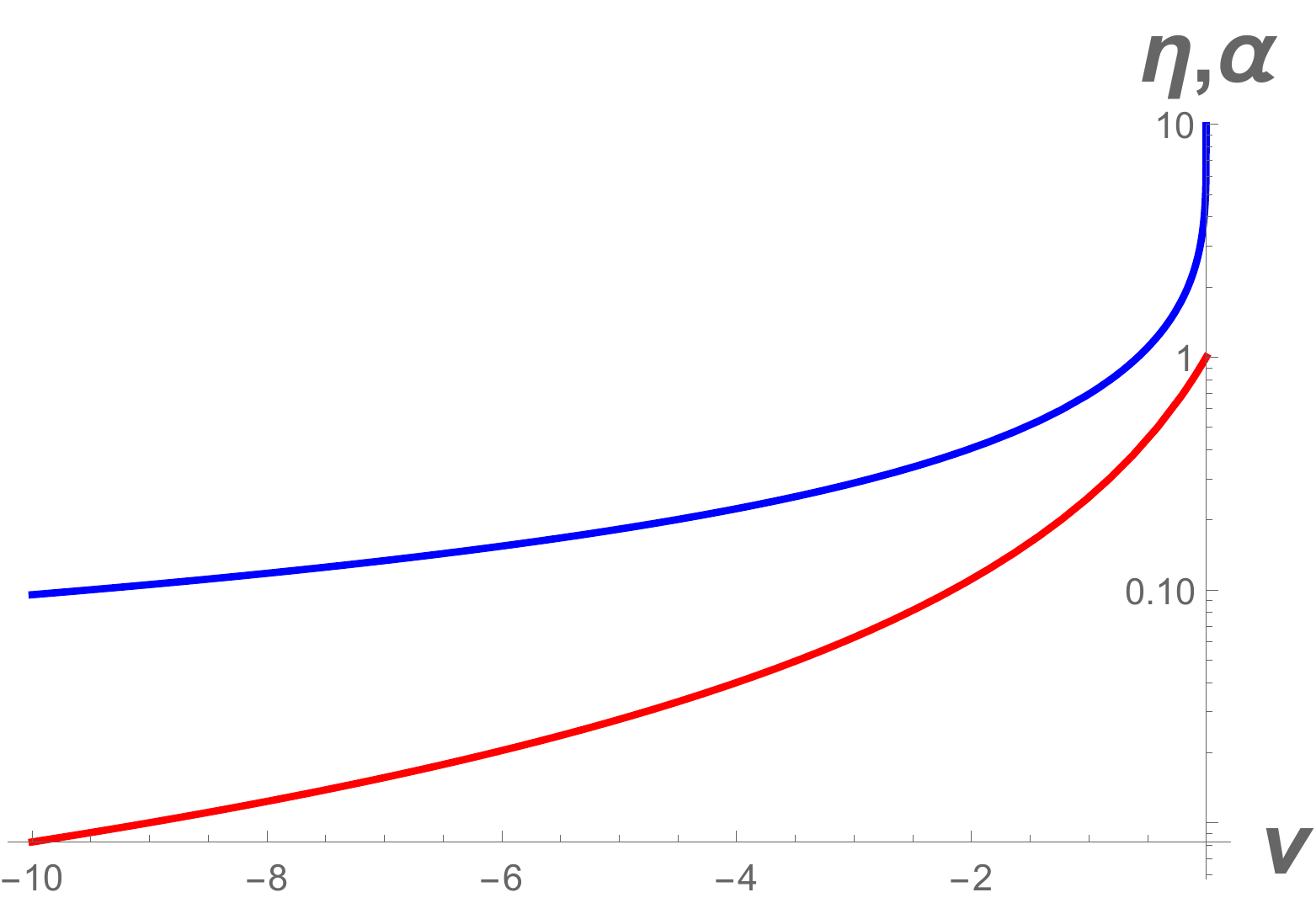} 
\caption{The rapidity, $|\eta(v)|$ of Eq.~(\ref{rapidity}), is illustrated in a log plot by the blue line that diverges (the mirror approaches the speed of light) at the $v_H = 0$ horizon, where $\kappa = 1$ units. The proper acceleration, $|\alpha(v)|$, of Eq.~(\ref{acc}), is illustrated in red, and asymptotically approaches $\kappa$ which has been set to $\kappa = 1$.    
}
\label{fig:RapidityPlot} 
\end{figure}

The rapidity, as defined through \cite{paper2},
\be f'(v) = e^{-2\eta(v)} \quad\rightarrow\quad \eta(v) = -\frac{1}{2}\ln \frac{d f(v)}{dv},\ee
as a function of advanced time, is
\be \eta(v)= \ln \left(1-\frac{1}{\kappa(v_H-v)+1}\right).\label{rapidity}\ee
The limit in the far past, $v\rightarrow -\infty$ is $\eta \rightarrow 0$;  the mirror is past asymptotically static.  The limit as advanced time $v \rightarrow v_H$, from below, is $\eta\rightarrow -\infty$.  The mirror rapidly travels left, off to the speed of light. The proper acceleration \cite{paper2}, through,
\be \alpha(v) = e^\eta(v) \frac{d\eta(v)}{dv},\ee
is a notably simple monotonic function of $v$,
\be \alpha(v) = -\frac{\kappa }{\left(\kappa(v_H-v)+1\right){}^2}.\label{acc}\ee
In the limit that advanced time approaches the horizon, 
\be \lim_{v\rightarrow v_H} \alpha = -\kappa.\ee
This character is the key defining dynamic trait of the mirror, i.e.  the mirror becomes uniformly accelerating in the far future.   Both the magnitudes of rapidity and acceleration are plotted in Fig.~(\ref{fig:RapidityPlot}) as a function of advanced time, $-\infty < v < v_H=0$.
\section{Entanglement Entropy}\label{sec:EE}
 \subsection*{Black Hole as 1-D}
  In order to get a handle on a non-thermal black hole like ERN, consider the more familiar case of any thermal black hole emitting radiation.  There is a good reason to believe that such a black hole, like a 3-D Schwarzschild, has information emission more similar to a 1-D channel than a surface in 3-D space \cite{Bekenstein:2001tj}. Namely, because information flow for all systems at a certain temperature is 
  \be \dot{S} \sim P/T, \ee
  in all space dimensions.  The power dependence on the temperature of the system in question can tell us about the information transmission. In a 1-D system (like the thermal moving mirror where energy flux is $\pi T^2/12$), the power is 
  \be P \sim T^2. \label{1Dpower} \ee
  However for a 3-D black body surface, using the Stefan-Boltzmann law, we have 
  \be P \sim A T^4, \label{SB} \ee
  which seemingly deviates from Eq.~(\ref{1Dpower}).  However, for a Schwarzschild black hole, temperature $T \sim M^{-1}$, we see the area is $A\sim M^2$ or $A \sim T^{-2}$.  Substituting this into Eq.~(\ref{SB}) we retrieve Eq.~(\ref{1Dpower}). Therefore, a black hole acts as a 1-D information channel transmitter.  This helps justify partially but precisely the use of the 1-D moving mirror as a blueprint for investigating entanglement information during collapse to a thermal black hole. But can these ideas be extended to a non-thermal system, like the ERN solution? 

\subsection*{CFT entropy with a boundary}
Extension starts with the entropy for a system in 1+1 dimensions in conformal field theory with size $L$ and UV cut-off $\epsilon$ \cite{geoentropy, Cal}:
\be S = \frac{1}{6} \ln \frac{L}{\epsilon}.\ee
The size of the system is the trajectory of the mirror,
\be L \equiv p(u) - p(u_0), \ee
The smearing is asymmetric, $\epsilon^2 \equiv \epsilon_{p}\epsilon_{p_0}$, and the relationship between the smearing and the motion is
\be \epsilon_{p} = p'(u)\epsilon_u,  \ee
and likewise for the other edge of the system,
\be \epsilon_{p_0} = p'(u_0)\epsilon_{u_0}. \ee
We apply this setup by treating the case of a static and arbitrarily moving boundaries as separate cases. 
\subsubsection*{Static Mirror vs. Arbitrary Motion}
\begin{figure}[ht]
\begin{center}
\includegraphics[width=1.5in]{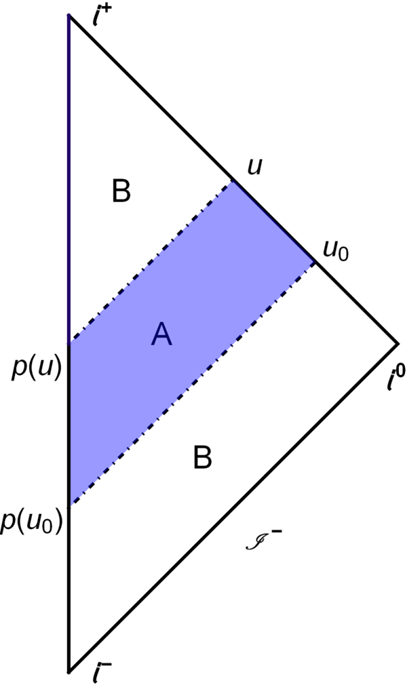}
\caption{\label{fig:static} A static mirror and the entangled subsystems.  Notice $B$ is in the past and future.} 
\end{center}
\end{figure} 
In the static mirror example, we see that the size is simply $L = u - u_0$, and smearing is $\epsilon^2 = \epsilon_u \epsilon_{u_0}$.  Therefore the vacuum entropy is just
\be
S_{\textrm{vac}} = \frac{1}{12}\ln\frac{(u-u_0)^2}{\epsilon_u\epsilon_{u_0}}.
\ee
The subsystems for a static mirror are plotted in Fig.~(\ref{fig:static}).  This system is trivial: no particles, no energy, no spectra.  
However, for an arbitrary moving mirror, we can see that the entropy is given by:
\be S_{\textrm{bare}} = \frac{1}{12}\ln\frac{[p(u)-p(u_0)]^2}{p'(u)p'(u_0)\epsilon_u\epsilon_{u_0}},\ee
which is also explicitly dependent on the smearing.  Arbitrary motion is non-trivial, i.e. here there is vacuum radiation.  
\subsubsection*{Renormalized Entropy}
We can remove the dependence on smearing by use of a renormalized entropy, found by simple subtraction of the vacuum entropy from the bare entropy,
\be S_{\textrm{ren}}= S_{\textrm{bare}} - S_{\textrm{vac}}, 
\ee
which gives a smearing independent entropy,
\be S_\textrm{ren} = \frac{1}{12}\ln\frac{[p(u)-p(u_0)]^2}{p'(u)p'(u_0)(u-u_0)^2}.\ee
This is still not wholly satisfactory, because in most moving mirror cases of interest (initial collapse has zero particle production), we will want to compare with the asymptotically static past: the vacuum state where the field modes look like natural Minkowski plane waves.  This implies we should specialize and extend the region $A$ by sending $u_0$ back to the asymptotic past.
\subsubsection*{Past-Future Entanglement}
\begin{figure}[ht]
\begin{center}

\includegraphics[width=2.3in,height=2.3in,keepaspectratio]{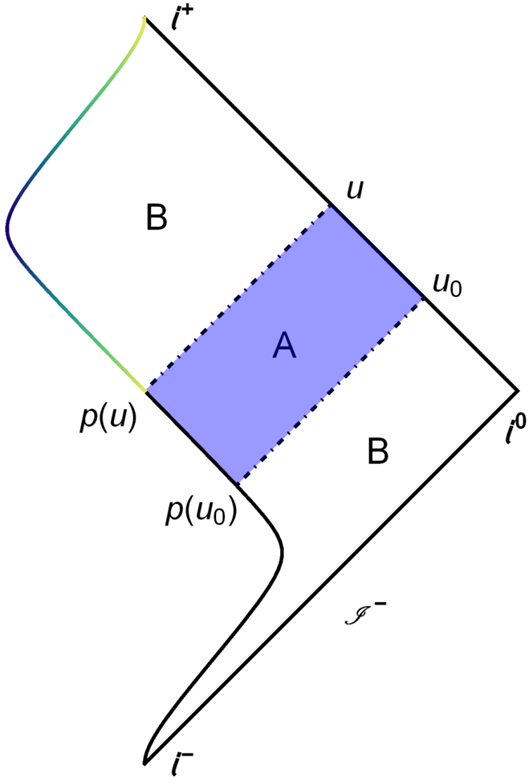} \;\;\;
\includegraphics[width=2.3in,height=2.3in,keepaspectratio]{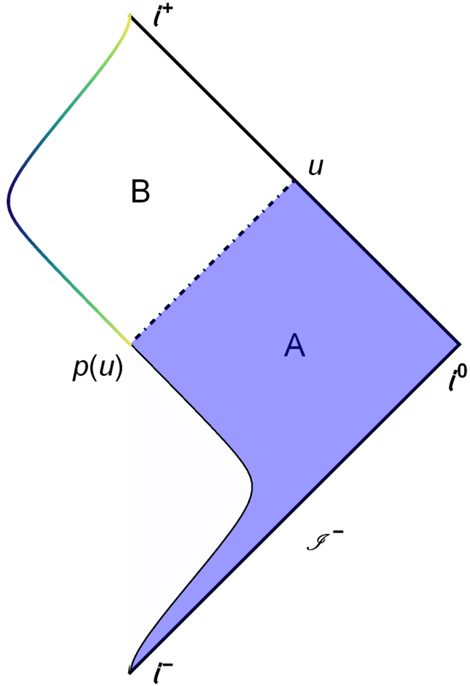} 
\caption{\label{past} An arbitrarily moving mirror system at some point in time $u$,  described by $p(u)$, can have the sub-system $A$ extended to include the entire past by taking the limit $u_0 \rightarrow -\infty$. This is the key assumption needed to identify rapidity with entanglement entropy.} 
\end{center}
\end{figure}

A significant simplification occurs when expanding sub-system $A$ to include the entire past.  This is done by taking the limit $u_0 \rightarrow -\infty$, as illustrated in Fig.~(\ref{past}), so that $S_\textrm{ren} \rightarrow S(u)$, where $S(u)$ is now dependent only on $u$,  
\be S(u) = -\frac{1}{12}\ln p'(u) \ee
The dynamical meaning of this equality is clarified by transcendental inversion to spacetime coordinates, $(t,x)$, from null coordinates $(u,v)$, through 
\be p'(u) = \frac{1+\dot{x}(t)}{1-\dot{x}(t)},\ee
yielding \cite{paper3,yeomentropy,signatures}
\be S(t) = -\frac{1}{6} \tanh^{-1} \dot{x}(t), \ee
from which it is seen that since rapidity, $\eta(t) \equiv \tanh^{-1} \dot{x}(t)$, then, independent of coordinates:
\be S = -\frac{1}{6}\eta. \ee
Therefore entanglement entropy is rapidity.  In hindsight, one could have guessed this was the case because of the additive nature of both entropy and rapidity.  So it is clear, that the faster the mirror moves away, the greater the entanglement.  A zero velocity mirror demonstrates zero entanglement between the entrie past and entire future subsystems.  A rapidly approaching mirror (positive rapidity) gives negative entropy. The simplicity of this relationship makes it a practical tool in the context of mirrors and may be useful in other dynamic systems where entanglement entropy arises.   

\subsection*{Thermodynamic Entanglement}
How is the entanglement entropy related to the thermodynamic entropy?  Consider for a moment the eternal thermal mirror of Carlitz-Willey \cite{CW}. 
Its entanglement entropy $S_{EE} = -\eta/6$ is found\footnote{Transcendental inversion for spacetime coordinates are tractable for the trajectory \cite{spin, timedep}.}, with $\eta(u) = -\kappa u/2$ as:
\begin{equation}
S_{EE}(u)=\frac{\kappa u}{12}=\frac{\pi T u}{6}.\label{ETentropy}
\end{equation}
Its thermodynamic entropy density is
\begin{equation}
\overline{S}_T\equiv \frac{\partial F}{\partial T}=\frac{\pi T}{6}\ ,
\end{equation} 
where $F= (768\pi M^2)^{-1} = \kappa^2/(48\pi) =\pi T^2/12$ for the eternal thermal mirror, using $\kappa \equiv (4M)^{-1}$. Defining an entanglement entropy density, $\overline{S}_{EE}$, with respect to null time $u$,
\begin{equation}
\overline{S}_{EE}\equiv\frac{S_{EE}}{u}=\frac{\pi T}{6},
\end{equation}
one sees that,
\begin{equation}
\overline{S}_{EE}=\overline{S}_T,\label{ETdensity}
\end{equation}
i.e.\ entanglement entropy density is always thermodynamic entropy density for the eternally thermal moving mirror system. 

The ERN mirror, with $\eta$ given by Eq.~(\ref{rapidity}), is decidedly not thermodynamically entangled.  Therefore, the particles will not be distributed according to Planck, nor will the stress tensor exhibit a global steady state constant.  However, we can still use the concept of entanglement entropy, as the mirror still has a rapidity.  Namely, its immediatley clear that the rapidity diverges and information loss is present as $v\rightarrow v_H$, see Fig.~(\ref{fig:RapidityPlot}).  Visually tracing the modes can be done in Figs. (\ref{fig:ERNpenrose}) and (\ref{fig:PenrosePlot}).
\section{Stress Tensor and Total Energy}\label{sec:stress}
\subsection*{Stress Tensor}
Balbinot et al. \cite{Balbinot:2007kr} found that the stress energy tensor of the non-extreme RN black hole stays regular in the extreme limit and smoothly transitions to that of non-extreme black holes. Contrary to previous studies, it was shown that the expectation values of the quantum stress energy tensor for our massless scalar field living on a 1+1 dimensional ERN black hole background are indeed completely regular on the horizon \cite{Balbinot:2004jx}. These results provide confidence to utilize the stress tensor in the moving mirror model, integrated to the horizon, which we use to find total evaporation energy.  The stress tensor is the Schwarzian derivative of Eq.~(\ref{f(v)}), \cite{horizonless}  
\be F(v)= \frac{1}{24\pi}\{f(v),v\}f'(v)^{-2},\ee
which yields,
\be F(v) = \frac{\kappa^2}{6\pi}\frac{\left(\kappa(v_H-v)\right){}^3}{  \left(\kappa(  v_H-v)+1\right){}^6}.\label{F(v)}\ee
The maximum flux, $F_m= (384\pi M^2)^{-1}$, is half as much as the usual Hawking flux for a Schwarzschild black hole, $F_H = (768\pi M^2)^{-1}$, comparing equal mass stars (`charged' vs uncharged).  Even though charged black holes are colder than neutral black holes, recall that the ERN cannot be characterized by temperature \cite{Anderson:2000pg,LRS}, so it is not actually cooler, in spite of the intuitive result, $F_H = 2F_m$.

\begin{figure}[ht]
\centering 
\includegraphics[width=3.2in]{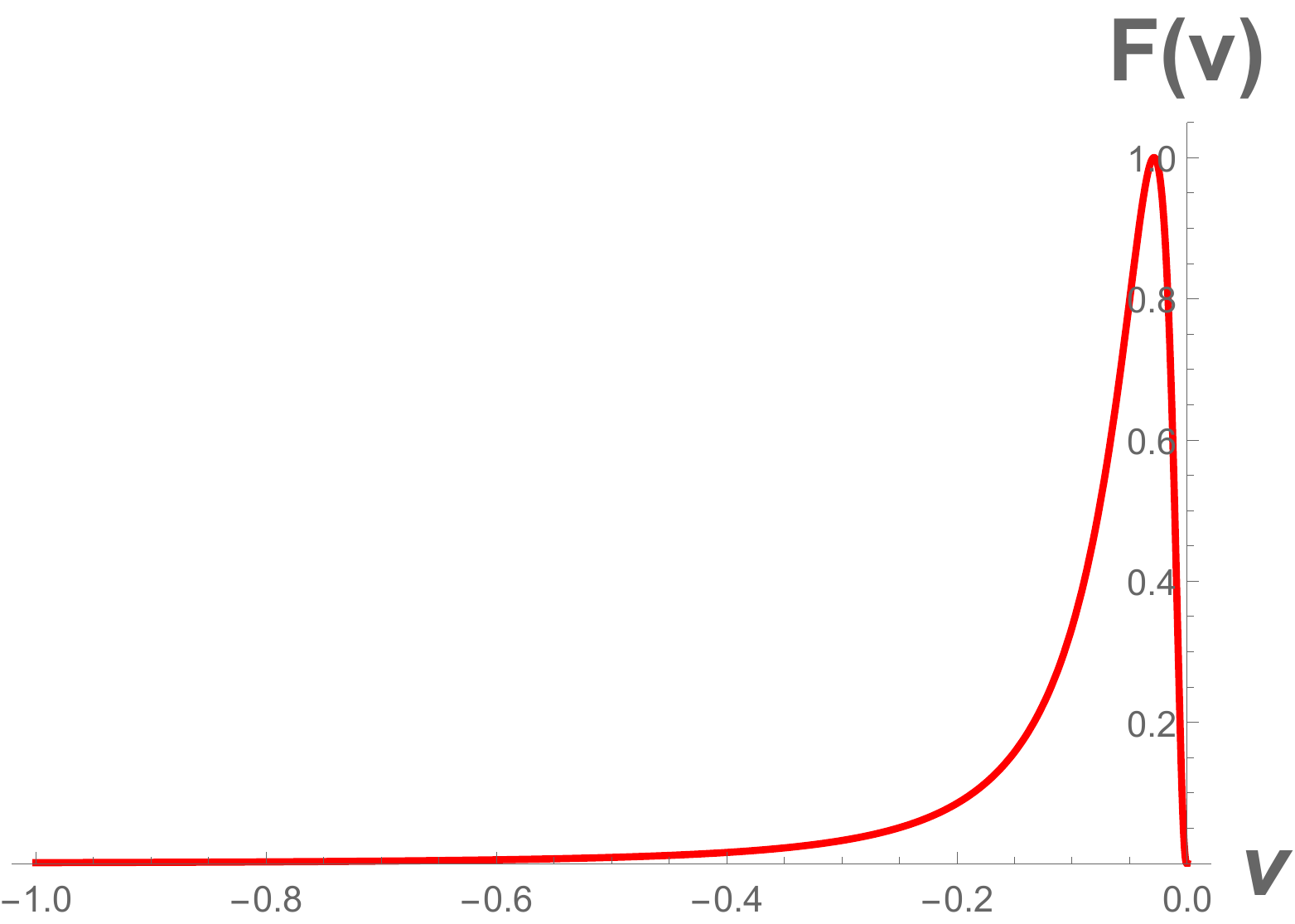} \caption{ The energy flux, $F(v)$, Eq.~(\ref{F(v)}), radiated by the ERN moving mirror is illustrated in advanced time with horizon at $v=v_H =0$. In units $\kappa = \sqrt{384\pi}$, so that the maximum flux is normalized to $F_m=1$. The time is $v=-1/\kappa = -2M$, when the flux is maximum.       
}
\label{fig:FluxPlot2} 
\end{figure}

\begin{figure}[ht]
\centering 
\includegraphics[width=3.2in]{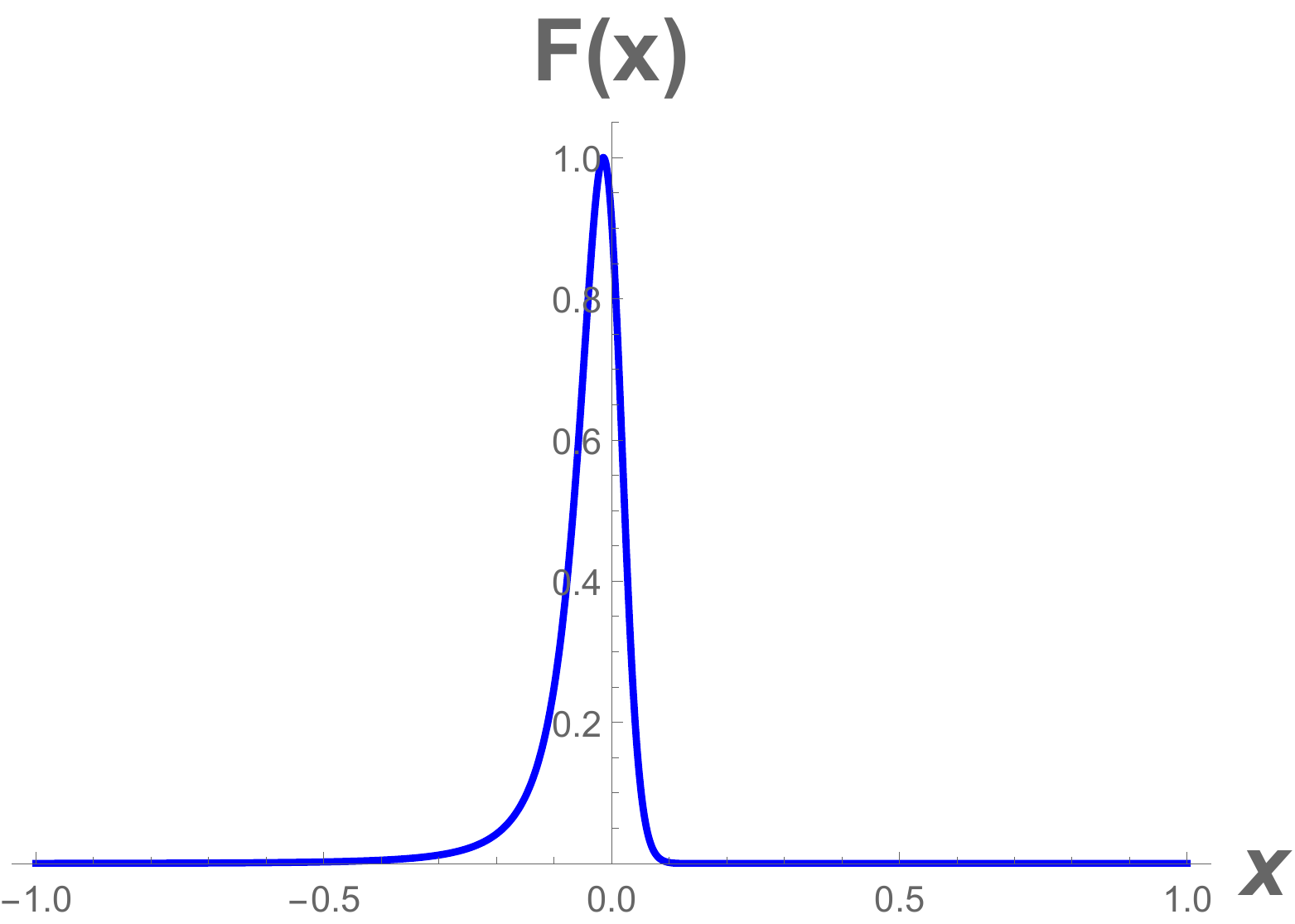} \caption{ The energy flux, Eq.~(\ref{F(x)}) radiated by the ERN moving mirror is illustrated in space, $F(x)$, by the blue line, in units where $\kappa = \sqrt{384\pi}$, so that the maximum flux is normalized to $F_m=1$.  The location this happens at is $x = -1/(2\kappa) = -M$.  Recall that the mirror moves to the left, so that $x=+\infty$ is asymptotic past times, $t = -\infty$.  By plotting in space, $x$, we have extended the coordinate variable to all $\infty$, in contrast to Fig.~(\ref{fig:FluxPlot2}).      
}
\label{fig:FluxPlot} 
\end{figure}

It is worth mentioning there is no transient negative energy flux.  Recall that advanced time ranges from $-\infty < v \leq v_H$.  This range is illustrated in Fig.~(\ref{fig:FluxPlot2}), and one observes the flux is always positive. This is a relatively unusual trait among solved moving mirrors, as all currently (2020) solved drifting mirrors are accompanied by negative energy flux (e.g. \cite{Good:2016HUANG}).  Negative energy flux is a requirement for mirrors that are asymptotically static or drifting \cite{GLW}. To look across space, we can express the flux in spacetime coordinates via 
\be F(x) = \frac{1}{12\pi}\frac{t'''(t'^2 -1)-3t't''^2}{(t'-1)^4(t'+1)^2},\ee
where primes are derivatives w.r.t. space coordinate $x$.  One gets
\be F(x) = \frac{4\kappa^2}{3\pi} \frac{W\left(\frac{1}{2}e^{-\kappa x}\right)^3}{\left(2 W\left(\frac{1}{2}e^{-\kappa x}\right)+1\right)^6}.\label{F(x)}\ee
This form is more intuitive, since there is no bound in space, i.e. the mirror covers the space, $+\infty > x > -\infty$, i.e. it travels the entire Minkowski grid, but starting from $x=+\infty$, moving left.  The flux, $F(x)$, is plotted in Fig.~(\ref{fig:FluxPlot}). 
\subsection*{Total Energy}
The total energy observed by our observer at $\mathscr{I}^+_R$ (see e.g. \cite{Ritus}) is worth calculating, even though the mirror accelerates forever, using Eq.~(\ref{F(v)}) and Eq.~(\ref{f(v)}),
\be E = \int_{-\infty}^{v_H} F(v) \frac{d f(v)}{dv} dv, \ee
where we integrate over delayed time, $du$, necessitating the introduction of the Jacobian $f'(v)$, and only up to the horizon, $v_H$.  The spacetime result, Eq.~(\ref{F(x)}) and Eq.~(\ref{t(x)}), is also convenient via numerical integration,
\be E = \int_{+\infty}^{-\infty} F(x) \left(\frac{dt(x)}{dx}-1\right) dx,\ee
because the bounds are infinite, remembering the mirror moves to the left starting at $x=+\infty$. The total energy result is analytic, 
\be E = \frac{\hbar \kappa}{36\pi}.\label{energy}\ee
While positive flux is accordant, finite energy is more than welcome.  All known asymptotically drifting mirrors emit a finite amount of energy (see e.g. \cite{GTC,universe,MG15}). In contrast, most known null-mirrors, i.e. those that attain the speed of light; produce an infinite amount of energy. This mirror maintains finite non-zero energy and asymptotic light speed with its eternal asymptotic constant acceleration.\footnote{One unsolved mirror that has finite energy but accelerates to infinite proper acceleration is investigated in \cite{paper3}.} 

The evaporation energy, $E$, should at least be less than the total mass, $M$, of the black hole: $E<M$.  In which case, one has an lower bound  on the mass,
\be M > \frac{1}{72\pi M}, \quad\quad \rightarrow \quad\quad M > 0.067\;m_P.\ee
In light of the regime of applicability of semi-classical gravity, this constraint is not violated.  The finite energy prevents the formation of a naked singularity by the emission of neutral scalar particles.  

\section{Spectrum and Particle Count}\label{sec:particles}
\begin{figure}[ht]
\centering 
\includegraphics[width=3.2in]{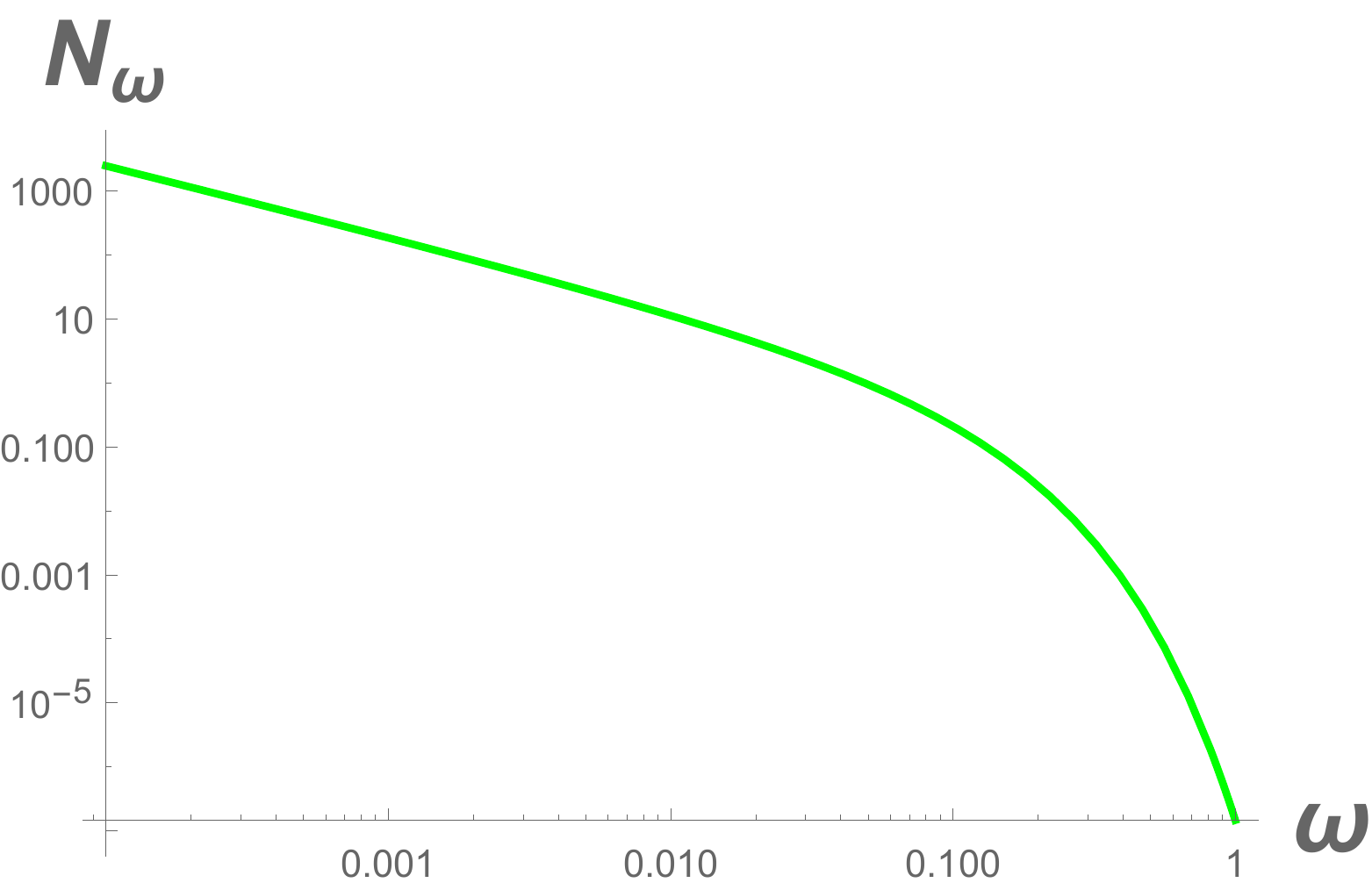} \caption{ The particle spectrum, Eq.~(\ref{N(w)}), in a log-log plot, demonstrating an infrared divergence, for $\omega\rightarrow 0$. This signals infinite total particle count due to the soft particles at $\omega =0$.       
}
\label{fig:ParticleSpecPlot} 
\end{figure}

As we have mentioned, LRS \cite{LRS} are well aware that the quantum radiation emitted by an incipient extreme black hole is not characterized by a temperature at any time during collapse, which has been backed up by Anderson-Hiscock-Taylor who found that macroscopic {\it{zero temperature}} black hole solutions do not exist \cite{Anderson:2000pg}.  Our results confirm these findings, and we extend this program of investigation further by solving for and analyzing the non-thermal spectrum. The beta Bogolubov coefficient can be found via,
\be \beta_{\omega\omega'} = \frac{-1}{4\pi\sqrt{\omega\omega'}}\int_{-\infty}^{v_H} dv e^{-i \omega' v -i \omega f(v)}\left(\omega f'(v)-\omega'\right),\ee
by setting the horizon $v_H=0$ for convenience and definiteness, (horizon position will not affect the spectrum because of complex conjugation).  The result is
\be \beta_{\omega\omega'}= \frac{-i e^{-\frac{\pi  \omega }{\kappa }}}{\pi \kappa}  \sqrt{\frac{\omega '}{\omega_p }} \left(\frac{\omega }{\omega_p }\right)^{\frac{i \omega }{\kappa }} K_{a}\left(\frac{2}{\kappa }\sqrt{\omega \omega_p}\right),\ee
where $\omega_p \equiv \omega + \omega'$, and $a \equiv \frac{2 i \omega }{\kappa }+1$. On our way to obtain the spectrum, the complex conjugate is taken,
\be |\beta_{\omega\omega'}|^2 = \frac{e^{-\frac{2 \pi  \omega }{\kappa }} \omega ' }{\pi ^2 \kappa ^2 \omega _p}\left|K_a\left(\frac{2}{\kappa }\sqrt{\omega  \omega _p}\right)\right|^2,\label{ERNbeta2}\ee
which is the particle count per mode squared.  The spectrum, which is the main result of this paper, is then,
\be N_\omega = \int_0^\infty  |\beta_{\omega\omega'}|^2 d\omega',\label{N(w)}\ee
plotted in Fig.~(\ref{fig:ParticleSpecPlot}).
An infrared divergence, signaling soft particles \cite{soft}, and infinite total particle count,
\be N =\int_0^\infty  |\beta_{\omega\omega'}|^2 d\omega d\omega',\ee
is illustrated.  The divergence of global particle count at zero frequency is unsolved in the moving mirror model for non-asymptotically static mirrors \cite{paper1,walkerdavies}.  At this juncture, it is good to test our intuition, that the particles (even though infinite) carry the energy \cite{Walker}.  A consistency check is warranted on the result, Eq.~(\ref{ERNbeta2}).  Summing up all the energies of each particle should yield the total finite energy:
\be E = \int_0^\infty \omega |\beta_{\omega\omega'}|^2 d\omega d\omega' =\frac{\hbar\kappa}{36\pi}.\ee
Indeed, this is not hard to check.  The result confirms the answer, Eq.~(\ref{energy}), obtained by use of the stress tensor.  
\subsection*{Time Evolution of Particle Production}
\begin{figure}[ht]
\centering 
\includegraphics[width=3.2in]{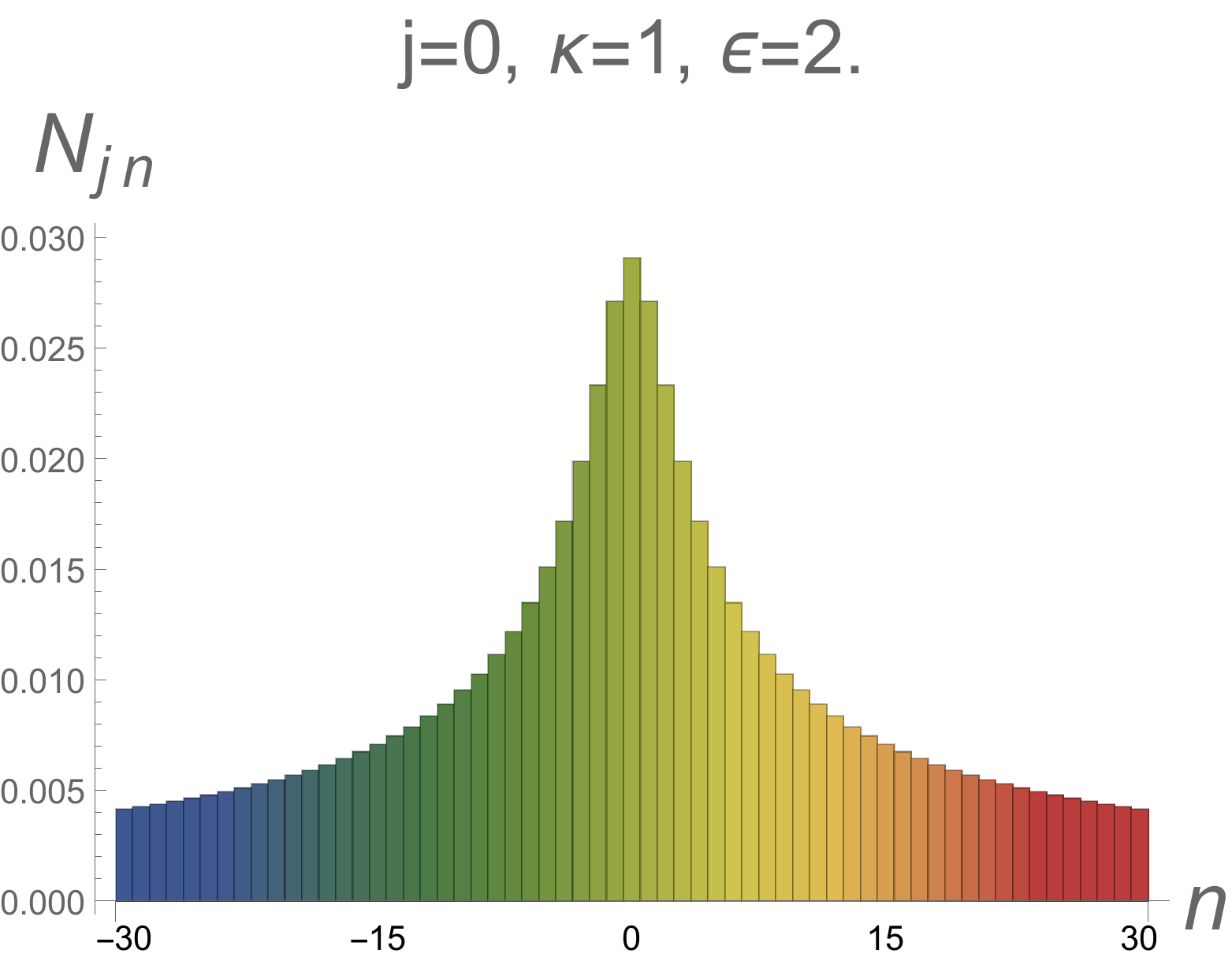} \caption{ The particle count in time, via wave packet localization.  The detector is set with $\epsilon = 2$. The scale of the system is $\kappa =1$ and the frequency bin is in the lowest possible $j=0$ value, where most of the particle production occurs, and finer resolution in time is possible. This emission includes the `phantom radiation' of \cite{LRS}.  It is symmetric in delayed time, $u$, centered around time bin $n=0$.    
}
\label{fig:TimePlot} 
\end{figure}

The time dependence of particle creation can be computed via wavepacket analysis suggested by Hawking \cite{Hawking}, and elaborated by others \cite{MPA, GLW}. Wave packet localization, particularly via othornormal and complete sets in the moving mirror model is described in detail in several works \cite{timedep,Good:2016HUANG,wpU}.  A plot of the particle creation in time is give in Fig.~(\ref{fig:TimePlot}).

\begin{figure}[ht]
\centering 
\includegraphics[width=3.2in]{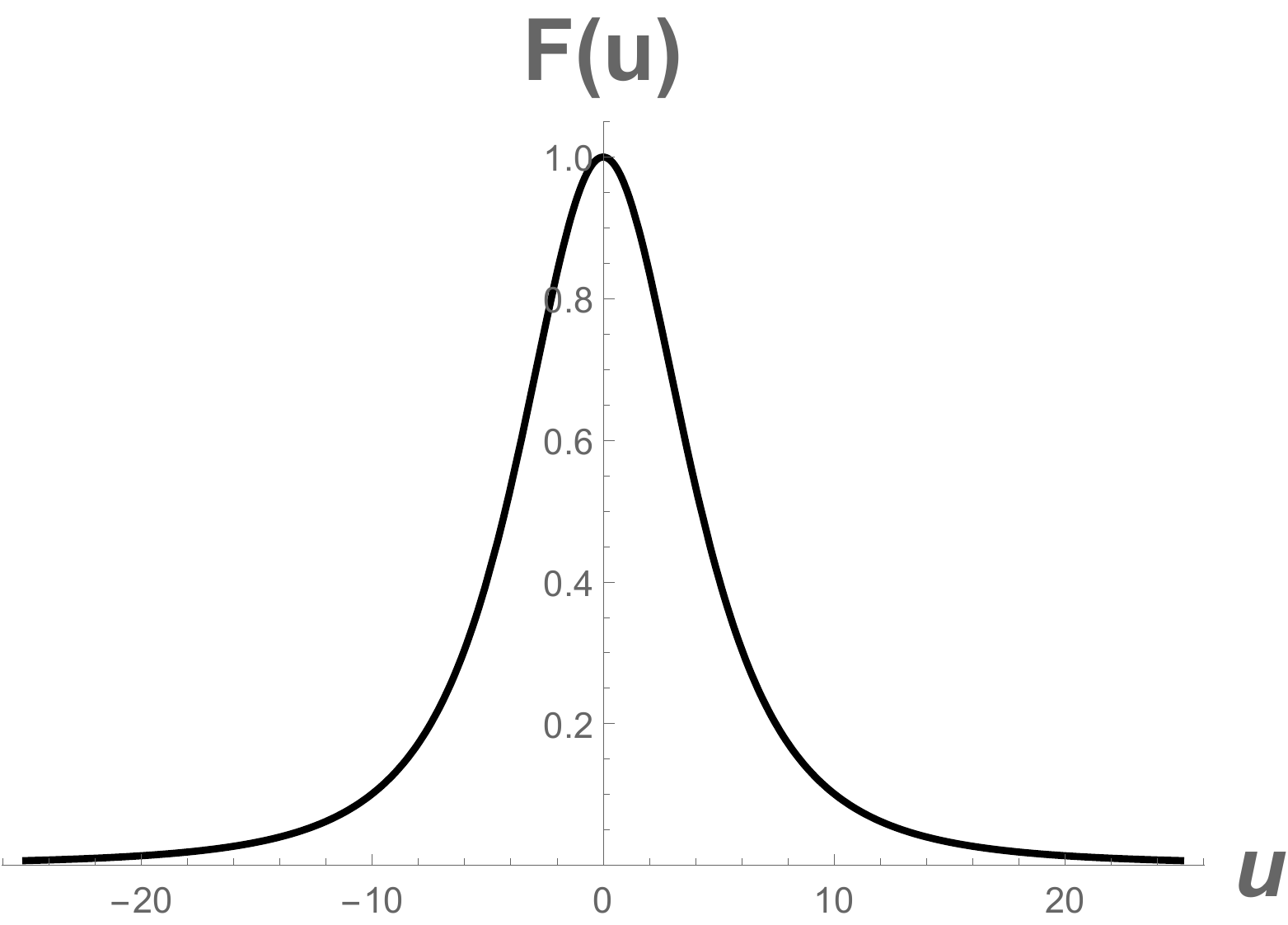} \caption{ The energy flux, $F(u)$, Eq.~(\ref{F(u)}), radiated by the ERN moving mirror is illustrated in delayed time $u$ with horizon at $v=v_H =0$. In units $\kappa = \sqrt{384\pi}$, so that the maximum flux is normalized to $F_m=1$. The time is $u=0$, when the flux is maximum. This plot demonstrates symmetry in delayed time, reflecting the symmetry of particle production in delayed time bins, $n$, of Fig.~(\ref{fig:TimePlot}).        
}
\label{fig:FluxPlot3} 
\end{figure}

The symmetric distribution of particle production looks at odds with the asymmetric energy flux production of $F(x)$ and $F(v)$ in Figs.~(\ref{fig:FluxPlot2}) and (\ref{fig:FluxPlot}), respectively.  The energy is carried by the particles, as we have seen, but is it carried asymmetrically in time?  The time bins $n$ correspond to delayed time $u = t-x$.  This compels a third look at the flux, now as $F(u)$, to confirm suspicion that the asymmetry is a coordinate artifact.  This can be done numerically, by an inverse function technique applied to $f(v)$, Eq.~(\ref{f(v)}). Using asymptotic acceleration units, the result for the flux, is
\be F(u) = -\frac{p(u)^3}{6\pi(p(u)-1)^6},\label{F(u)}\ee
plotted in Fig.~(\ref{fig:FluxPlot3}).  The result is symmetric in delayed time $u$, and the total integrated flux,
\be E = \int_{-\infty}^{+\infty} F(u) du,\ee
is in agreement with the total energy, Eq.~(\ref{energy}).  The symmetry here confirms the energy is carried symmetrically in delayed time, $u$, as also observed by a detector at $\mathscr{I}_R^+$, receiving the particles, delayed in time, $u$.

\subsection*{Uniform Acceleration Spectra}
It behooves us to compare the spectra of uniform acceleration with the ERN spectra, Eq.~(\ref{ERNbeta2}).  The uniformly accelerated mirror, $\alpha = -\kappa$,
\be p(u) = \frac{u}{1+\kappa u},\ee
with early-time horizon positioned at $u_H = -\kappa^{-1}$, has a spectrum that is solved via,
\be \beta_{\omega\omega'} = \frac{1}{4\pi \sqrt{\omega\omega'}}\int_{-\kappa^{-1}}^\infty du e^{-i u \omega -i \omega' p(u)}\left(\omega'p'(u)-\omega\right),\ee
via a simple substitution, $1+\kappa u = \kappa X$, so that the range of integration can be extended over $X$.  A heaviside theta function assists in further extension to cover $(-\infty,+\infty)$ integration. The result is
\be \beta_{\omega\omega'} = \frac{i e^{ \frac{i (\omega-\omega')}{\kappa} }  } {\pi\kappa} K_1\left(\frac{2}{\kappa} \sqrt{\omega \omega'}\right).\ee
The phase vanishes upon complex conjugation,
\be |\beta_{\omega\omega'}|^2 = \frac{1}{\pi^2\kappa^2} \left| K_1\left(\frac{2}{\kappa} \sqrt{\omega  \omega '}\right)\right|^2.\label{UAbeta2}\ee
This is the spectrum of a uniformly accelerated mirror, distinctly non-thermal \cite{Davies1,Davies2,B&D}.  It is straightforward to compare Eq.~(\ref{UAbeta2}) with Eq.~(\ref{ERNbeta2}) at late times to leading order, using the high frequency limit, $\omega'\gg\omega$, and one finds identical spectra. 
\section{Discussion}\label{sec:DISC}

A freely falling frame is a locally inertial frame that detects no gravitational force.  The ERN solution has zero surface gravity.  This time independence is reflected in the moving mirror. 

Consider that a free-falling observer on Earth registers on their accelerometer, zero acceleration.  To someone standing on Earth, the falling detector appears to accelerate downward to the Earth  (where the standing observer feels the force of the ground upward on their feet). This is the equivalent proper acceleration felt in the situation of the mirror. The mirror asymptotically feels a constant force, like that of the floor on the feet of an observer standing on Earth.  

The ERN zero gravity and asymptotically uniform acceleration mirror share this deep equivalence, but there are immediate and interesting limitations and extensions that are nevertheless present.

\subsection*{Limitations}
Our approach has been to keep everything as simple as possible by utilizing the moving mirror model, but the need to accept complications when clearly unavoidable is obviously present w.r.t. numerous physical aspects. While conceptually clear, the extreme RN moving mirror model is ultimately extremely limited, for instance:  
\begin{itemize}
    \item Grey-body factors:  the spectra will be altered by back-scattering \cite{Fabbri, LRS} and the nontrivial metric coefficients of spacetime.  These affects are not accounted for in the moving mirror model.
    \item Vacuum polarization: divergences in vacuum polarization terms \cite{Balbinot:2004jx} may alter the quantum stress tensor behavior in the curved spacetime contexts, however, in the ERN case, this appears to work in the model's favor by regularizing results.
    \item Dimensionality and spherical symmetry:  In order to appropriately generalize to 3+1 dimensions, the powerful simplification due to spherical symmetry apparently severally restricts utilization of the model; however some aspects have carry-over \cite{Tony}, e.g. zero surface gravity.
    \end{itemize}
An additional important caveat here is the nature of the fine-tuned collapse.  LRS assumed that models can be found in which collapse leads to a black hole with maximum charge \cite{LRS}.  The fine-tuning required to produce extreme solutions makes the question of their existence non-trivial \cite{noexist}.  The fine-tuning exists because of the extreme sensitivity to effects of backreaction \cite{Marolf:2010nd} of the quantum radiation on the metric or the mirror.  Nevertheless, as we have shown, interesting information can be found by assuming collapse occurs and deducing the consequences.  
\subsection*{Extensions}
Immediate extensions are possible.  First, a calculation of the variance.  The asymptotic value of the two-point function does not tend to zero which is the thermal emission value.  This could shed light on the information contained in the stress energy above and beyond what is apparent from the tensor alone, i.e. exponential fast damping exhibited by non-extreme collapse vs. power law damping exhibited by the ERN solution \cite{LRS}.  

Secondly, and particularly interesting, is a better understanding of the $\omega = 0$ singularity responsible for the soft particle divergence \cite{soft}. Removal by radiative corrections analogous to bremsstrahlung cross section cancellation may be possible \cite{LRS}, accounting for momentum transfer from field to mirror (recoil).  It may also be possible to simply regularize the spectrum by the uniform acceleration contribution, since only soft particles are characterized by these colors.  

Third, a concrete demonstration of the (in)stability \cite{noexist, Anderson:1995fw, Hod:2013eea} of the moving mirror dynamics would be useful, as early collapse emission could cause asymptotic runaway acceleration or asymptotic zero acceleration, pushing the system out of gravitational-electrical equilibrium. Different dynamics can lead to a calculation of radiation for the non-extreme RN mirror.  We will demonstrate this spectrum in a separate paper.

\section{Conclusions}\label{sec:conc}

We have solved the ERN black hole - moving mirror correspondence\footnote{The Schwarzschild mirror is solved here: \cite{Good:2016MRB,Good:2016LECOSPA,MG14one,MG14two}.}.  In particular we have made progress on several fronts: 
\begin{itemize}
\item Trajectory:  We have identified the world-line of the ERN moving mirror.  The choice of mirror trajectory is not arbitrary but corresponds uniquely to the outside geometry of the ERN.  Measurement of particle production at early times can violate the no-hair theorem, demonstrating details specific to the form of collapse. 
\item Entanglement: We have derived a simple formula, $-6S=\eta$, for the entanglement between the entire past and entire future subsystems, resulting in an information-dynamics relationship, characterized by rapidity, which is additive like entropy.

    \item Finite energy: Despite reaching asymptotic light speed, with never ending uniform acceleration, the evaporation process finishes.  This is signaled by the computed finite energy, $E = (72\pi M)^{-1}$.
    \item Spectra: We have a full evaluation of the spectra for all times.  Detection, even if possible, of arbitrarily low frequency soft particles registering non-zero detector probability, will not allow one to determine the spectra of collapse. Having the full spectra means we can possibly screen our detector from the non-informative (no information-carrying) flux of soft particles. 
    \item Time evolved particle count:  The evolution of particle creation is demonstrably non-thermal, and in fact peaks in out-of-equilibrium early time.  The non-steady rate explicitly verifies a non-Plankcian distribution. The particle count is damped in the far-past and far-future.  The ERN black hole releases information quickly (before late times) and the half-way point of collapse is reminiscent of \cite{Hayden} where the black hole acts like a quantum information mirror.    
\end{itemize}
     With respect to how quantized matter fields can preserve the CCC, we have prevented the formation of a naked singularity analog in the mirror system, maintaining cosmic censorship, by demonstrating the limiting energy emission and late-time zero energy flux.  The ERN does not lose mass at late times.   However, what about the no-hair theorem? The particle spectrum, valid for all-times, depends critically on the evolution of collapse but the late-time behavior, in fact, does not depend on the details of early collapse, actually preserving the late-time no-hair theorem. You lose hair when you are old.

     

\appendix

\acknowledgments 

MG thanks Paul Anderson, Eric Linder, Yen Chin Ong, and Daniele Malafarina for stimulating discussions. Funding from state-targeted program ``Center of Excellence for Fundamental and Applied Physics" (BR05236454) by the Ministry of Education and Science of the Republic of Kazakhstan is acknowledged. MG is also funded by the ORAU FY2018-SGP-1-STMM Faculty Development Competitive Research Grant No. 090118FD5350 at Nazarbayev University. 

\newpage

\onecolumngrid
\appendix


\end{document}